\documentclass[aps,showkeys,superscriptaddress,preprint,tightenlines]{revtex4}
\usepackage{hyperref}
\usepackage{times}
\usepackage{amsmath}
\usepackage{graphicx}
\usepackage{epsfig}
\usepackage{dcolumn}
\usepackage{bm}
\usepackage{color}
\usepackage{longtable}
\usepackage{array}
\usepackage{multirow}

\newcommand{\BR}{{\cal B}}

\newcommand{\pip}{\pi^+}
\newcommand{\pim}{\pi^-}
\newcommand{\piz}{\pi^0}

\newcommand{\etap}{\eta^{\prime}}

\newcommand{\jpsi}{J/\psi}

\newcommand{\EE}{e^+e^-}
\newcommand{\MM}{\mu^+\mu^-}

\newcommand{\ra}{\rightarrow}

\newcommand{\beq}{\begin{equation}}
\newcommand{\eeq}{\end{equation}}
\newcommand{\bitm}{\begin{itemize}}
\newcommand{\eitm}{\end{itemize}}

\newcommand{\pio}{\pi^0}

\newcommand{\ar}{\rightarrow}

\newcommand{\azz}{a_{0}(980)}
\newcommand{\fz}{f_{0}(980)}

\newcommand{\chico}{\chi_{c1}}

\begin{document}

\title{Light Meson Physics at \mbox{BESIII}}
\author{Shuang-shi Fang}
 \email{fangss@ihep.ac.cn}
 \affiliation{Institute of High Energy Physics, Chinese Academy of Sciences,
 Beijing 100049, China}
 \affiliation{University of Chinese Academy of Sciences, Beijing 100049, China}

\begin{abstract}

Studies of light meson decays are important tools to perform precision tests of the effective field theories,  determine transition form factors, and test fundamental symmetries.  
With very high statistics data samples, the \mbox{BESIII} experiment provides a unique laboratory for light meson studies and  is contributing significantly to a variety of these investigations.  
A brief review of  recent progress in light meson decay studied at the \mbox{BESIII} experiment, including  detailed studies of  common decay dynamics,  searches for rare/forbidden decays and  new particles,  is presented. Finally, together with descriptions of different experimental techniques,   prospects for future studies of light mesons are  discussed in some detail.  

\end{abstract}


\maketitle

\section{Introduction}
\label{Sec:intro}

The discoveries of light mesons and detailed studies of their decays have played  crucial roles in the development of our understanding of elementary particle physics.  In the case of weak interactions, important insights were gained from kaon and pion decays, such as  the observation of CP violation and the validation of V-A structure of the theory. In addition, the discovery of strangeness  inspired the SU(3) flavor symmetry,  which, in turn,  gave the birth to  the  quark model picture of the underlying structure of observed particles.  To date,  about seven decades  since the discovery of the first light mesons (the pion and kaon), studies of  light meson decays continue to  provide opportunities for a  variety of physics at low energy scales,  including precision tests of  effective  field theories, investigations of the quark structure of the light mesons,  tests of the fundamental symmetries, and searches for new particles.

The \mbox{BESIII}  experiment~\cite{Ablikim:2009aa}    collected the world's largest samples of $1.3\times 10^9$ $J/\psi$ events~\cite{Ablikim:2016fal} and $4.5\times 10^8$ $\psi(3686)$ events~\cite{Ablikim:2017wyh}  produced directly from  $e^+e$ annihilation in 2009 and 2012. 
Due to the high production rates of light mesons in the charmonium decays, these data,  in combination with the excellent performance of the detector, offer unprecedented opportunities to explore the light meson decays.  Moreover, the \mbox{BESIII} data sample of $e^+e^-$ annihilation events at energies between 2.0 and 3.08 GeV with an integrated luminosity of $~650$ pb$^{-1}$ allows for explorations of  properties of the light vector mesons, in particular the vector strangeonium states.

\section{Precision tests of QCD at low energies }

At high energies, QCD serves as a reliable and useful theory, whereas at low energies non–perturbative QCD calculations are usually performed by an effective field theory called Chiral Perturbation Theory (ChPT).  High quality and precise measurements of low-energy  hadronic
processes  are necessary in order to verify the systematic ChPT expansion. 
Thus,  studies of light meson decays are  important guides to our understanding of  how QCD works in the non-perturbative regime.

\subsection{Light quark mass ratios in $\eta/\eta^\prime\rightarrow 3\pi$ decays}
The decay of the $\eta$ meson into 3$\pi$ violates  isospin symmetry, which is related to  the difference of light-quark 
masses,  $m_u \neq m_d$.  Therefore the decay of $\eta\rightarrow 3\pi$  offers a unique way to determine the quark mass ratio
 $Q^2\equiv (m_s^2-{\hat m}^2)/(m_d^2-m_u^2)$ (where ${\hat m} = \frac{1}{2}(m_d + m_u)$ ).  Extensive theoretical studies have been performed within the framework of combined ChPT and dispersion theory~\cite{Gasser:1984pr,Bijnens:2007pr,Schneider:2010hs,Kampf:2011wr,Guo:2015zqa, Colangelo:2018jxw}.

In addition to the recent results from the WASA-at-COSY~\cite{Adlarson:2014aks} and KLOE-2~\cite{Anastasi:2016cdz} experiments,  \mbox{BESIII} reported a Dalitz plot analysis of $\eta\rightarrow 3\pi$ decays~\cite{Ablikim:2015cmz}. The measured matrix elements are in agreement with the most precise KLOE-2 determination and theoretical predictions.
Taking experimental results as input, 
two dedicated analyses  presented the results, $Q= 22.0 \pm 0.7$~\cite{Colangelo:2016jmc}  and  $Q = 21.6 \pm 1.1$~\cite{Guo:2016wsi} ,  respectively.  In the near future, the study of
$\eta \ra \pi^{+}\pi^{-}\pi^0$ and $\eta\ra\pi^0\pi^0\pi^0$ decays  at \mbox{BESIII} will provide an  independent check of these results by directly fitting to  differing theoretical models.

Historically, the $\eta^\prime\rightarrow \pi^+\pi^-\pi^0$ decay was considered to proceed via $\pi^0-\eta$ mixing~\cite{Gross:1979ur}, which offered the possibility of comparable strength  $u$-$d$ quark mass difference
from the branching fraction ratio of $r={{\BR}(\etap\ra\pi\pi\pi)}/{\mathcal{B}(\etap\ra\pi\pi\eta)}$.
However,  it was argued subsequently that  the  decay amplitudes are strongly
affected by the intermediate resonances~\cite{Borasoy:2006uv}, e.g., the $P$-wave contribution from $\eta^\prime\rightarrow\rho\pi$, and, thus,  $u-d$ quark mass difference could not be extracted in such a simple way.

\begin{figure}[hbtp]
\begin{center}
  \includegraphics[width=9.cm,height=9.cm]{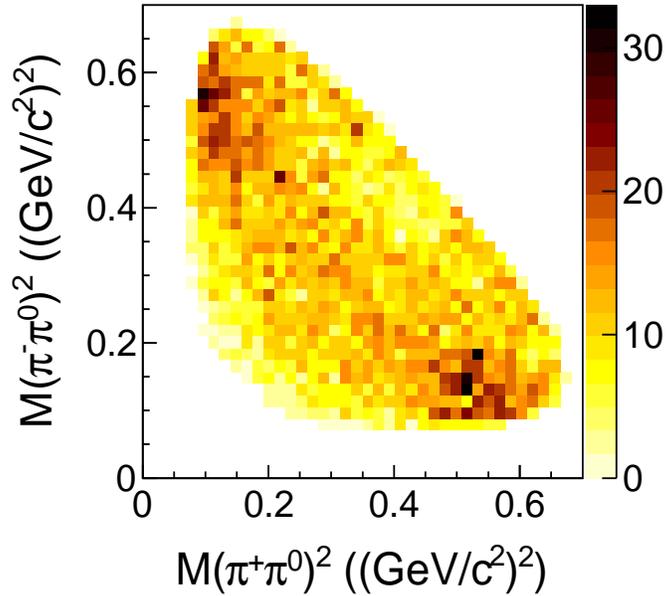}
     \caption{Dalitz plot of $M^2(\pi^+\pi^0)$ versus $M^2(\pi^-\pi^0)$ for the
$\etap\ra\pip\pim\pio$ decay, where the two clear clusters correspond to the $\eta^\prime\to\rho^\mp\pi^\pm$ decay ~\cite{Ablikim:2016frj}.
      }
  \label{m3pidalitz}
\end{center}
\end{figure}

In addition to the first observation of  $\etap\ra\rho^{\pm}\pi^{\mp}$ (Fig.~\ref{m3pidalitz}) by \mbox{BESIII}~\cite{Ablikim:2016frj},  
the resonant $\pi$-$\pi$
$S$-wave, interpreted as the broad $f_0(500)$, is also expected to play an essential role in $\eta^\prime\to\pi^+\pi^-\pi^0$ decays.  
The contribution of  $f_0(500)$  provides a
reasonable explanation for the negative slope parameter of the
Dalitz plot of $\etap\ra\pio\pio\pio$~\cite{Ablikim:2015cmz}.
Due to limited statistics,  it has been impossible to differentiate
between $S$ and $D$ waves; larger event samples are  crucial
for carrying out amplitude analyses of these processes.
Several theory groups have expressed interest to describe the decay using a
dispersive approach.  These improved theoretical studies along with more precise experimental measurements of  $\eta/\eta^\prime\rightarrow 3\pi$ decays from a variety of experiments are expected to improve the accuracy of the quark mass ratio.

\subsection{Cusp effect in $\eta^\prime\rightarrow \pi^0\pi^0\eta$ decays}

In addition to the test of the effective theoretical models, common to all $\eta^\prime\rightarrow \pi\pi\eta$ decays,  the neutral decay  $\eta^\prime\rightarrow \pi^0\pi^0\eta$ 
 also allows us to  examine the cusp effect, $i.e.$, an abrupt change in the $\pi^0\pi^0$ invariant mass distribution as it crosses the 2$m_{\pi^+}$ threshold.
 An accurate measurement of the cusp  effect may  enable a determination of the S-wave pion–pion scattering lengths to high precision.

For  $\eta^\prime\rightarrow\pi^+\pi^-\eta$, \mbox{BESIII}  results~\cite{Ablikim:2017kp}  are  not  consistent well with  theoretical predictions based on the chiral unitary approach~\cite{Borasoy:2005du}.   The discrepancies show up as about four standard deviations on some of the parameters that are used to describe the Dalitz plot distribution.
 In the case of  $\etap\ra\pio\pio\eta$, the results are in general consistent with  theoretical
predictions within the uncertainties and  the latest results reported by the A2 experiment~\cite{Adlarson:2017wlz}.
Due to  the limited statistics,  the present results are not precise enough to firmly establish
isospin violation and additional effects, {\it e.g.}, radiative corrections~\cite{Kubis:2009sb} and the $\pi^+/\pi^0$
mass difference should be considered in the future experimental and theoretical studies.

At \mbox{BESIII} search for the cusp in $\etap\ra\eta\pi^0\pi^0$ performed by inspecting  the $\pi^0\pi^0$ mass spectrum
close to $\pip\pim$ mass threshold~\cite{Ablikim:2017kp}, revealed no statistically significant effect.  
From an experimental perspective, the available high-statistics of 10 billion $J/\psi$ events at \mbox{BESIII}  is expected to increase the  $\eta^\prime$ decay event sample by nearly an order of 
magnitude. These additional data coupled with the incorporation of recent dispersive theoretical analyses~\cite{Isken:2017dkw} make  an investigation of the cusp effect in this channel very promising.

\subsection{Box anomaly in $\eta/\eta^\prime\rightarrow\gamma\pi^+\pi^-$ decay}

In the Vector Meson Dominance (VMD) model, the main contribution to the decay $\eta^\prime\rightarrow\gamma\pi^+\pi^-$ comes from $\eta^\prime\rightarrow\gamma\rho$.
However, a  significant deviation in the dipion distribution between the theory predictions and data is observed, and this may be attributable  to the Wess-Zumino-Witten box anomaly~\cite{box-anomaly,box-anomaly2}. The previous measurements~\cite{Althoff:1984jq,Aihara:1986sp,Albrecht:1987ed,Bityukov:1990db,Benayoun:1992ty,Abele:1997yi}  give sometimes opposite conclusions on the presence
of the box anomaly term.

\begin{figure}[hbtp]
\begin{center}
  \includegraphics[width=9.cm,height=7.5cm]{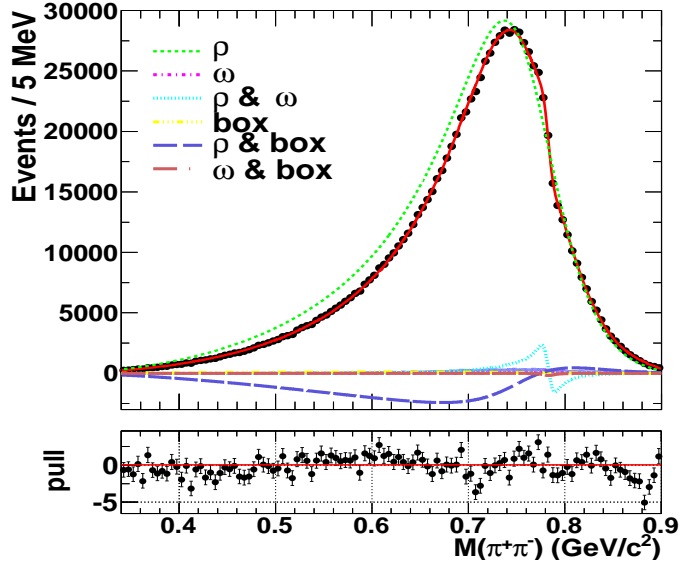}
     \caption{The results of the model-dependent fits to $M(\pip\pim)$ with $\rho^0$-$\omega$-box anomaly~\cite{Ablikim:2017fll}.
      }

  \label{etap-invidata}
\end{center}
\end{figure}

Recently, a precision \mbox{BESIII} study of $\eta^\prime\rightarrow\gamma\pi^+\pi^-$ ~\cite{Ablikim:2017fll} found, for the first time, that  a fit that only included the components of $\rho$ and $\omega$ and their  interference failed to describe the data; a significant additional contribution, either the box anomaly or a $\rho(1450)$ component, is found to be necessary, as indicated in Fig.~\ref{etap-invidata}, to provide a good description of data.
In this case, the influence  of the box anomaly phenomenon , $i.e.$, the presence of  a well-defined contact term is still awaiting a definite and unambiguous demonstration.

The large and clean $\eta/\eta^\prime$ sample produced in $J/\psi$ decays at \mbox{BESIII} is expected to  promote the study of  $\eta/\eta^\prime\rightarrow\gamma\pi^+\pi^-$  into an unprecedented precision era.  Along with a recently proposed model-independent approach~\cite{Stollenwerk:2011zz},
 a combined analysis of $\eta/\eta^\prime\rightarrow\gamma\pi^+\pi^-$ may present a consistent picture for the dynamics of these two decays. 

\subsection[]{Test of higher-order ChPT with \boldmath $\eta/\etap\to\gamma\gamma\pio$ and  $\etap\to \gamma\gamma\eta$ decays }

The decays $\eta/\eta^\prime\to\gamma\gamma\pi^0$ are of particular interest for tests  of  ChPT at the two-loop level.  Since light vector mesons play a critical role in these models, the dynamical role of the vector mesons has to be  systematically included in the context of either the VMD or Nambu-Jona-Lasinio model to reach a deeper understanding of these decays.

The $\eta\to\gamma\gamma\pi^0$ decay has been measured by many experiments~\cite{pdg2000}. Of interest is that  the branching fraction of 
 $\eta\to\gamma\gamma\pi^0$, ${(8.4\pm 2.7\pm 1.4)}\times 10^{-5}$~\cite{DiMicco:2005stk},  as reported by KLOE is approximately a factor of three lower than that from the A2 experiment~\cite{Nefkens:2014zlt}. 
Experimentally, both the  $\eta^\prime\rightarrow\gamma\gamma\pi^0$~\cite{Ablikim:2016tuo} and  $\eta^\prime\rightarrow\gamma\gamma\eta$~\cite{Ablikim:2019wsb} decays were studied at \mbox{BESIII}. 
The measured branching fractions are in agreement with a recent theoretical calculation based on the Linear sigma model with VMD couplings~\cite{Escribano:2018cwg}.
It was also found that the di-photon invariant mass dependence of the partial decay widths differs in shape from predictions of  the different theoretical models~\cite{Escribano:2018cwg}. Thus a precision measurement  of the di-photon  mass spectrum would be a more sensitive tool for testing the reliability of theoretical calculations than just measurements of the branching fraction. 
In this case, an updated  measurement for these double radiative decays using the full $J/\psi$ sample  at \mbox{BESIII} will provide an opportunity to have a combined analysis that will distinguish between  different theoretical models.

\subsection{Transition form factors of light mesons}

The  $\eta/\eta^\prime\to\gamma l^+l^- (l=e,\mu)$ Dalitz decays,  where the lepton pair is formed by internal conversion of an intermediate virtual photon and 
the decay rates are modified by the electromagnetic structure arising at the vertex of the transition, are of special interest.  Deviations of measured quantity  from their QED predictions are usually described in terms of  a
 timelike transition form factor, which, in addition of being an important  probe  into the meson's structure~\cite{Landsberg:1986fd},  has an important role in the evaluation of the hadronic light-by-light  contribution to the muon anomalous magnetic moment (see a nice review~\cite{Aoyama:2020ynm} for details).

\begin{figure}[hbtp]
\begin{center}
 \includegraphics[width=9.cm,height=7cm]{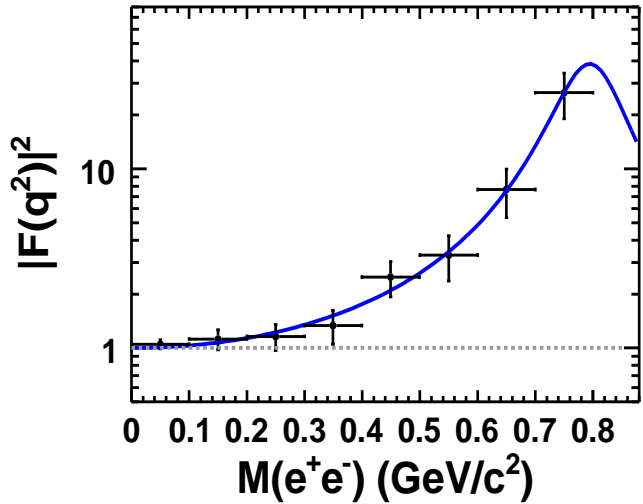}
  \caption{ Fit to the single pole form factor  $|F(q^2)|^2$ , where $q^2$ is  the square of the $e^+e^-$ invariant mass~\cite{Ablikim:2015wnx}. }
  \label{etap:fit}
           \end{center}
\end{figure}

In contrast to SND and WASA experiment's studies of  $\eta\to\gamma l^+l^-$ \cite{snd,wasa}, \mbox{BESIII} has a unique advantage in the study of Dalitz decays of both $\eta$ and $\eta^\prime$ due to their high production rate in $J/\psi$ radiative and hadronic decays.  \mbox{BESIII}  reported the first measurement of the $e^+e^-$ invariant-mass distribution for  $\eta^\prime\rightarrow \gamma e^+e^-$~\cite{Ablikim:2015wnx}. It was  found that the single-pole parameterization provides a good description of data as illustrated in Fig.~\ref{etap:fit}. The corresponding  slope parameter,
$b_{\eta'}=(1.56\pm0.19)$~GeV$^{-2}$, is in agreement with the predictions from different theoretical models~\cite{Bramon:1981sw,Ametller:1983ec,Ametller:1991jv,Hanhart:2013vba} and a 1979 previous measurement of 
 $\eta'\to \gamma \mu^+\mu^-$~\cite{Dzhelyadin:1979za}.

 The decays  $\eta/\eta^\prime\rightarrow l^+l^- l^+l^-$ address decays via two off–shell
photons and indicate whether double vector meson dominance is realized in
nature.  To date, only the decay $\eta\rightarrow e^+e^- e^+e^-$ was observed at KLOE~\cite{KLOE2:2011aa}.  
The corresponding form factor has neither been
measured in the timelike nor  the spacelike region. 
In accordance with the theoretical investigation in Ref.\cite{Petri:2010ea}  predicted decay rates of $\eta^\prime\rightarrow e^+e^-e^+e^-$ of the order of $10^{-4}$,   hundred of events are expected to be
 observed at \mbox{BESIII} and significant progress could be made to test the latest theoretical prediction of $2.1\times 10^{-6}$ \cite{Escribano:2015vjz} based on a data-driven approach.  

In addition, using the data sample  collected at a center-of-mass energy of 3.773 GeV by the \mbox{BESIII} experiment,  studies~\cite{Czerwinski:2012ry} show that the measurements of the spacelike transition form factors in the decay
$e^+e^-\rightarrow e^+e^- \pi^0(\eta,\eta^\prime)$ via $\gamma\gamma$ interactions  in the range of the transfer momentum $Q^2$ within $[0.3, 10]$ GeV/c$^2$ are feasible.  It is worth mentioning that more data
 samples at 3.773 GeV and higher  are planned for the \mbox{BESIII}. They will be useful for the spacelike transition form factor measurements that are  complementary to the data 
from other experiments   and uniquely cover the $Q^2$ range that is relevant to the hadronic
light-by-light correction for the evaluation of the muon anomaly moment.

\subsection{Cross channel effect in $\omega\rightarrow \pi^+\pi^-\pi^0$ decays}

The decay $\omega\to \pi^+\pi^-\pi^0$ is usually employed  to investigate the $\omega$ decay mechanism by comparing a high–statistics Dalitz plot density distribution  with theoretical predictions. 
In the dispersive theoretical framework~\cite{Niecknig:2012sj,Danilkin:2014cra}, the Dalitz plot distribution and integrated decay width are sensitive to  the so-called crossed-channel effect~\cite{Danilkin:2014cra}. However, prior to \mbox{BESIII}, 
no experimental $\omega\rightarrow \pi^+\pi^-\pi^0$ data of sufficient precision were available to compare with the predictions.

Due to  the high production rate of $\omega$  in $J/\psi$ hadronic decays,   
\mbox{BESIII} was able to perform a precision Dalitz plot analysis with a sample of  $2.6\times 10^5$ $\omega \rightarrow \pi^{+}\pi^{-}\pi^{0}$ events~\cite{Ablikim:2018yen}, which is about six times larger 
than the samples in the previous work~\cite{Adlarson:2016wkw}  by WASA-at-COSY.  
It was found that   the Dalitz plot distribution of data significantly  differs from  the pure $P$-wave phase space, and additional contributions from resonances and/or final-state interactions (FSI), are necessary.
However, with the present statistics, the experimental results are consistent with the theoretical predictions without the need for incorporating crossed-channel effects, which may indicate that  the crossed-channel effect contributions are overestimated in the dispersive calculations.
Thus, the investigation on this decay dynamics with higher precision by analyzing the full $J/\psi$ data sample is  needed to  clarify this issue. 

\section{Quark structure of light scalar mesons}

The nature of the light scalar mesons $f_0(500)$, $K^*_0(800)$, $a_0(980)$ and $f_0(980)$ has been a controversial issue for several decades.
Taking into account the observations in heavy meson decays,  the existence of these scalar mesons is no controversial, though $K^*_0(800)$ is still qualified as "needs confirmation" in the PDG listings~\cite{pdg2000}.  However, the properties of these scalar mesons cannot be understood as simple $q\bar{q}$ mesons, and  non-$q\bar{q}$ interpretations of the light scalar nonet are supported by a variety of theoretical approaches~\cite{Chen:2007xr,RuizdeElvira:2010cs,Hooft:2008we,Parganlija:2010fz}.

Compared to scattering experiments, $J/\psi$ decays provide a clean laboratory to explore these scalar states.  At BESII, a series of amplitude analyses were performed to study the 
scalar mesons decays into pseudoscalar meson pairs $\pi\pi$, $K\bar{K}$ and $\pi K$ in $J/\psi$  decays~\cite{Ablikim:2004qna,besf0,beskappa} that established  the existence of the $f_0(500)$ and $K^*(800)$.

\begin{figure}[htbp]
   \centering
   \vskip -0.2cm
  \includegraphics[width=12.0cm,height=8.0cm,angle=0]{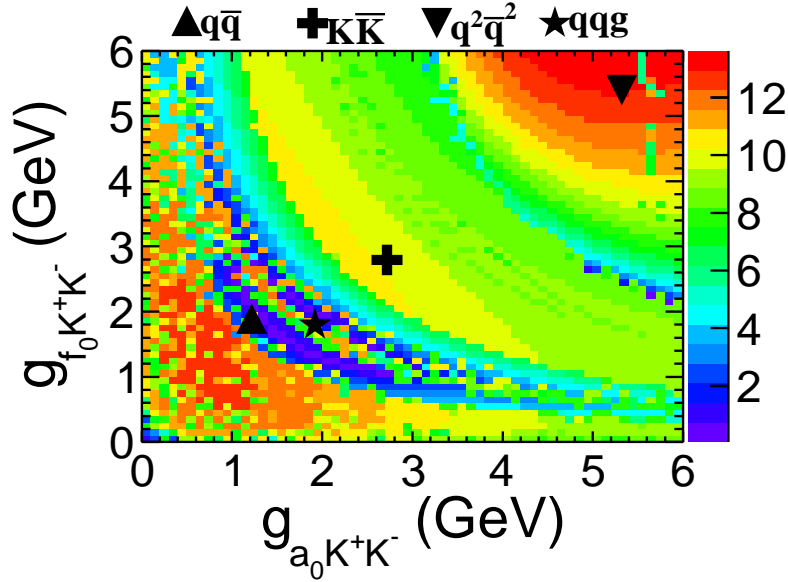}
   \vskip -0.1cm
   \hskip 0.0cm
   \caption{The statistical significance of the signal scanned in the two-dimensional space of $g_{a_{0}K^{+}K^{-}}$ and $g_{f_{0}K^{+}K^{-}}$~\cite{Ablikim:2018pik}, where 
   the markers indicate predictions from various illustrative theoretical models. The regions with
higher statistical significance indicate larger probability for the
emergence of the two coupling constants. }
   \label{signif}
 \end{figure}

 At \mbox{BESIII},  the $\azz$-$\fz$ mixing effect,  an essential approach for probing their nature, was observed for the first time in studies of $\jpsi\to\phi\eta\piz$  and $\chico\to\piz\pip\pim$ decays
 ~\cite{Ablikim:2018pik}.  The anomalous shape of $a_0(980)$ and the very narrow $f_0(980)$ peak produced by the  mixing effect was clearly seen in the  $\eta \pi^0$ and $\pi^+\pi^-$ mass spectra.  
The significance of the mixing effect  was then investigated as a function of  the two coupling constants, $g_{a_{0}K^{+}K^{-}}$ and $g_{f_{0}K^{+}K^{-}}$, and compared with 
 different models for the mesons' substructure, as shown in Fig. ~\ref{signif}. The results
 favor the  tetraquark model, although  other possibilities still can not be completely ruled out. 

In addition to their production via charmonium decays,  other processes can also be used to explore the properties of scalar mesons at \mbox{BESIII},  including light meson  and  charm meson decays. Examples are the prominent  $f_0(500)$ contribution in $\eta^\prime \rightarrow 3\pi$ decays~\cite{Ablikim:2016frj}, and the evident effects of $\azz$-$\fz$ mixing 
in an amplitude analysis of $D_s^+\to \pi^+ \pi^0\eta$~\cite{Ablikim:2019pit}.  Scalar mesons copiously produced in these decays 
are further evidences  that the \mbox{BESIII} experiment  is a unique facility for understanding  the controversial nature of these particles.

\section{Precision tests of  fundamental symmetries}

The $\eta$ and $\eta^\prime$ mesons are  eigenstates of $P$, $C$ and  $CP$ whose strong and electromagnetic decays are either anomalous or forbidden to  lowest order by  $P$, $C$, $CP$ and angular momentum conservation. Therefore, their decays  provide a unique  laboratory for testing the fundamental symmetries in flavor-conserving processes, which was extensively reviewed in Ref.\cite{Gan:2020aco}.

A straightforward way to test  these symmetries is to  search  for  $P$-  and $CP$-violating
$\eta/\eta^\prime$ decays into  two pions.  In the SM, the branching fractions for these modes  are very tiny~\cite{Jarlskog:2002zz}, but they may be enhanced by  $CP$  violation in  the extended  Higgs sector  of the electroweak theory~\cite{Jarlskog:1995gz}. 
 The high production rate for $\eta^\prime$ mesons in $J/\psi$ decays enabled \mbox{BESIII} to report the best experimental limit to date, $4.5 \times 10^{-4}$ , for $\mathcal{B}(\eta^\prime\rightarrow\pi^0\pi^0)$~\cite{Ablikim:2011vg} at the 90\% confidence level.
More recently, \mbox{BESIII}  made a search for the rare decay of
 $\eta^\prime\rightarrow 4\pi^0$ and reported the branching upper limit, ${\cal{B}}(\eta^\prime\to 4\pi^0)<3.8 \times 10^{-5}$ at the 90\% confidence level, for the first time~\cite{Ablikim:2019msz}.

Another interesting signal for possible CP violating mechanisms would be an asymmetry in the angle between the $\pi^+\pi^-$ and
$e^+e^-$ planes in the $\eta/\eta^\prime$ rest frame, where the  asymmetry  would be caused by  the interference between the usual $CP$ allowed magnetic  transition (driven by the chiral anomaly) and a $CP$ violating flavor-conserving electric dipole operator~\cite{Geng:2002ua}. The experimental bound on this asymmetry for $\eta\rightarrow \pi^+\pi^- e^+e^-$, $A_\phi = (-0.6\pm 3.1)\times 10^{-2}$~\cite{Ambrosino:2008cp},  from  the KLOE experiment is compatible with zero.  At \mbox{BESIII},  taking into account the measured  branching fraction  for $\eta^\prime\rightarrow \pi^+\pi^- e^+e^-$, $(2.11\pm0.12\pm
  0.15)\times10^{-3}$~\cite{Ablikim:2013wfg},  about $2\times 10^4$ events could be used to explore the CP violation using the full data sample of 10 billion $J/\psi$ events.
Most recently, the $\etap\ar\pip\pim\MM$  decay is  observed for the first time by the BESIII experiment~\cite{Ablikim:2020svz}.

Experimentally, $\eta/\eta^\prime\rightarrow l^+l^- \pi^0$ decays could be used to test  charge-conjugation invariance. In the SM, this process can proceed via  a two-virtual-photon exchange whereas one-photon-exchange violates  C-parity.   Within the framework of the VMD model, the most recent predictions~\cite{Escribano:2020rfs} for the branching fraction
are on the order of $10^{-9}$ for $\eta\rightarrow l^+l^- \pi^0$ and $10^{-10}$ for  $\eta^\prime\rightarrow l^+l^- \pi^0 (\eta)$. Thus, 
a significant enhancement of the branching fractions exceeding the two-photon model may be indicative of C violation.
With the available 10 billion $J/\psi$ events,  further improvement  for these rare decays will be achieved.

\section{Light quark vector mesons  in $e^+e^-$ annihilation}

Information on light vector meson decays has been obtained from $e^+e^-$ annihilations by, $e.g.$, the KLOE, SND, CMD-2, Babar and Belle experiments (see  Ref.~\cite{scattering} for a review), where the vector mesons are observed as the peaks in  the total cross section for the specific final states when the $e^+e^-$ center of mass energy is varied by tuning the beam energy or by the initial state radiation (ISR) process. With energy scan data in the $2.0-3.08$ GeV, \mbox{BESIII} can perform direct searches for light vector mesons, especially the poorly studied vector strangeonium states.

\begin{figure*}
\begin{center}
\includegraphics[width=8.6cm,height=5.8cm]{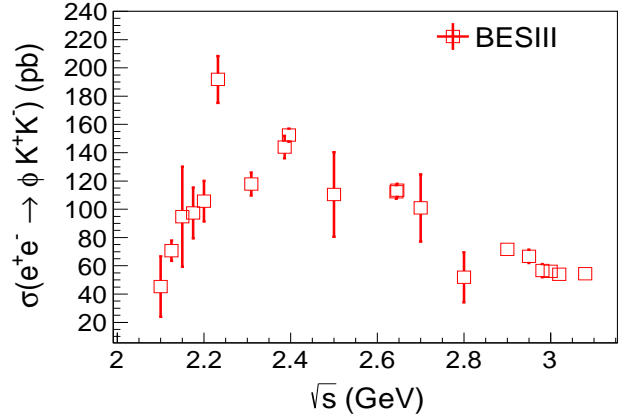}
\end{center}
\caption{The measured Born cross section of  $e^+e^-\to \phi K^+K^-$~\cite{KKKK}. }
\label{lineshape4K}
\end{figure*}

The $\phi(2170)$, previously referred to as the Y(2175), 
has been established by the BaBar~\cite{babar} and BES~\cite{bes} experiments, but its  measured mass and width remain controversial. 
There have been a number of different interpretations  for $\phi(2170)$, such as
a conventional $s\bar{s}$ state,  a QCD hybrid, tetraquark state, 
a $\Lambda\bar{\Lambda}$ bound state, 
or $\phi K\bar{K}$ resonance state.
The situation will not be clarified without further experimental data.
At \mbox{BESIII}, the line shapes of the cross sections for  a number of measured channels, including $e^+e^-\to K^+K^-$~\cite{besKK},   $e^+e^- \to K^+K^-\pi^0\pi^0$~\cite{KKpi0pi0}, and $e^+ e^- \to \phi\eta^\prime$~\cite{Ablikim:2020coo},  were measured  and a clear structure around 2.2 GeV  was evident in each of them. The measured widths and masses are consistent with 
 those  from $J/\psi\to\phi\pi^+\pi^-\eta$~\cite{Ablikim:2014pfc}, as summarized in  Table.~\ref{phi2170}.  Of interest is the process of  $e^+e^-\to K^+K^-K^+K^-$~\cite{KKKK} , and  its dominant 
 submode $e^+e^-\to \phi K^+K^-$. The  line shape for the latter is shown in Fig.~\ref{lineshape4K}.
In both cases, a very narrow enhancement at $\sqrt{s}= 2.232$~GeV is observed, 
which is very close to the $e^{+}e^{-} \to \Lambda \overline{\Lambda}$ 
production threshold.

 Another interesting possible strangeonium candidate is the $X(1750)$ observed in the photoproduction process~\cite{focus}, which was originally interpreted as the photoproduction mode of the $\phi(1680)$.  However, the recent simultaneous observation of the $\phi(1680)$ and $X(1750)$
 in  $\psi(2S)\to  K^+K^-\eta$ decays~\cite{Ablikim:2019xhx}   indicates that the $X(1750)$ is distinct from the  $\phi(1680)$ and possibly a strangeonium state. 
  
 The above examples demonstrate that \mbox{BESIII} is a powerful instrument for investigating the light vector mesons.   At present,  more studies, such as $e^+ e^- \to \phi\pi^+\pi^-$, $e^+ e^- \to \phi\eta$ and $J/\psi\to K^+K^-\eta $ are ongoing with 
 the aims of a deeper understanding of the nature of the $\phi(2170)$ and $X(1750)$, and searching for new strangeonium states.

\begin{table}[htpb]
\caption{\label{phi2170} Summary of mass and width of $\phi(2170)$ obtained from \mbox{BESIII}. }
\begin{tabular} {l |c|c}
\hline
Process & Mass (MeV/c$^2$) & Width (MeV) \\
\hline
$e^+e^-\to K^+K^-$~\cite{besKK} & $2239.2\pm7.1\pm11.3$ & $139.8\pm12.3\pm20.6$ \\
\hline
$e^+e^- \to K^+K^-\pi^0\pi^0$~\cite{KKpi0pi0} & $2126.5\pm16.8\pm12.4$ &$106.9\pm32.1\pm28.1$\\
\hline 
 $e^+e^-\to \phi \eta^\prime$~\cite{Ablikim:2020coo} & $2177.5\pm 4.8\pm 19.5$& $149.0\pm15.6\pm8.9$\\ 
 \hline
$J/\psi\to \phi \pi^+\pi^-\eta$~\cite{Ablikim:2014pfc} & $2200\pm 6\pm 5$&  $104\pm15\pm15$\\
\hline 

\end{tabular}
\end{table}

\section{Summary and prospects}
\label{Sec:Summary}

The light meson decays, as described above, provide a unique  opportunity to investigate many  aspects of particle physics at low energy, with the advantages of high production rates and the excellent performance of the detector. 
In addition to  improved accuracy on many of the measured properties of well known light meson decays, a series of first  observations, such as new decay modes of $\eta^\prime$, $a_0(980)-f_0(980)$ mixing as well as possibly new strangeonium states, were reported. These 
significant advances demonstrate that \mbox{BESIII} is playing a leading role in 
the study of light meson decays.

Despite this impressive progress, many  light meson  decays are still  unobserved and need to be explored.  At \mbox{BESIII},  $10^{10}$ $J/\psi$ events data are now available. This is  eight times larger than the subdata sample used in the present publications and  offers great additional opportunities for research in light meson decays, especially for pseudoscalar and vector mesons, with unprecedented precision. Moreover, 
\mbox{BESIII} expects to take an additional  20 fb$^{-1}$ of data at 3.773 GeV,  which will support investigations of the light meson physics  with different  ISR and two-photon production techniques,  such as the production of new vector mesons and measurements of the two photon width of the light scalar mesons.
In addition, different experimental techniques will  give access to previously unexplored regions of the electromagnetic transition form factors, allowing a quantitative connection between the timelike and the spacelike regions.

In general, together with the other high precision experiments, such as KLOE-2, A2, GlueX and BelleII,  these very abundant and clean event samples that are accumulated at \mbox{BESIII}  will bring the study of light meson decays into a precision era, and will definitely play an important role in the  developments of  chiral effective field theory and  Lattice QCD, and make significant contributions to understanding of hadron physics in the non–perturbative regime.

\section*{Acknowledgments}
The author appreciate Prof. S. L. Olsen for useful comments and suggestion,
This work is supported in part by  the National Natural Science Foundation of China (NSFC) under Contracts Nos. 11675184 and 11735014.
The author declares that we do not have any commercial or associative interest that represents a conflict of interest in connection with the work submitted.


\begin{thebibliography}{000}



\bibitem{Ablikim:2009aa}

Ablikim M, An ZH and  Bai JZ {\em et~al.} (\mbox{BESIII} Collaboration). 
Design and Construction of the \mbox{BESIII} Detector. 
 {\it Nucl Instrum Meth A} 2010;  {\bf 614}: 345-399.

\bibitem{Ablikim:2016fal} 

 Ablikim M, Achasov MN, Ai XC {\it et al.} Determination of the number of $J/\psi$ events with $J/\psi$ inclusive decays. {\it Chin Phys C} 2017;  {\bf 41}: 013001.
 
\bibitem{Ablikim:2017wyh} 

 Ablikim M, Achasov MN, Ahmed S {\it et al.} Determination of the number of $\psi(3686)$ events at BESIII. {\it Chin Phys C} 2018;  {\bf 42}: 023001.
 
\bibitem{Gasser:1984pr}
Gasser J and  Leutwyler H. $\eta\to 3\pi$ to one loop.
{\it  Nucl Phys B} 1985; {\bf250}: 539-560.

\bibitem{Bijnens:2007pr}
Bijnens J and  Ghorbani K.
$\eta\to 3\pi$ at two loops in Chiral Perturbation Theory
{\it J High Energy Phys} 2007;  {\bf 0711}: 030.

\bibitem{Schneider:2010hs}
Schneider SP, Kubis  B  and Ditsche C. 
Rescattering effects in $\eta\to 3\pi$ decays.
{\it J High Energy Phys} 2011;  {\bf 1102}: 028.


\bibitem{Guo:2015zqa}
Guo P, Danilkin IV and Schott D  {\em et~al.}
Three-body final state interaction in $\eta\to 3\pi$.
{\it Phys Rev D} 2015; {\bf 92}: 054016.


\bibitem{Colangelo:2018jxw}

Colangelo G, Lanz S and Leutwyler  H {\em et~al.}
Dispersive analysis of $\eta \rightarrow 3 \pi $.
{\it Eur. Phys. J. C} 2018; \textbf{78}:947


\bibitem{Kampf:2011wr}

Kampf K, Knecht M and Novotny J {\em et al.} 	
Analytical dispersive construction of $\eta\to 3\pi$ amplitude: first order in isospin breaking.
{\it Phys Rev D} 2011; {\bf 84}: 114015.


\bibitem{Adlarson:2014aks}
Adlarson P,  Augustyniak W and  Bardan W {\em et~al.} (WASA-at-COSY Collaboration). 	
Measurement of the $\eta\to\pi^+\pi^-\pi^0$ Dalitz plot distribution.
{\it Phys Rev C} 2014; {\bf 90}: 045207.

\bibitem{Anastasi:2016cdz}
Anastasi A, Babusci D and Bencivenni G {\em et~al.} (KLOE-2 Collaboration).
Precision measurement of the $\eta\rightarrow \pi^+\pi^-\pi^0$
 Dalitz plot distribution with the KLOE detector. 
{\it J High Energy Phys} 2016; {\bf 1605}: 019.

\bibitem{Ablikim:2015cmz}
Ablikim M, Achasov MN and Ai XC  {\em et~al.} (\mbox{BESIII} Collaboration). 
Measurement of the Matrix Elements for the Decays $\eta\to\pi^+\pi^-\pi^0$ and $\eta/\eta^\prime\to\pi^0\pi^0\pi^0$.
\newblock Phys  Rev  D 2015; {\bf 92}: 012014.

\bibitem{Colangelo:2016jmc}

Colangelo G, Lanz S and Leutwyler H {\em et al.} 
$\eta \to 3 \pi$: Study of the Dalitz plot and extraction of the quark mass ratio $Q$.
{\it Phys Rev Lett } 2017; {\bf 118}: 022001.


\bibitem{Guo:2016wsi}
Guo P, Danilkin IV and Fernandez-Ramirez C {\em et~al.} 
Three-body final state interaction in $\eta\to 3\pi$ updated.
 {\it Phys Lett B}  2016; {\bf 771}: 497-502.


\bibitem{Gross:1979ur}
 Gross DJ, Treiman SB and  Wilczek F.
  Light quark masses and isospin violation.
{\it Phys Rev  D} 1979; {\bf 19}: 2188-2214.

\bibitem{Borasoy:2006uv}
Borasoy B,  Meissner UG and Nissler R.
On the extraction of the quark mass ratio (m(d)- m(u)) / m(s) from $\Gamma(\eta^\prime \to \pi^0 \pi^+ \pi^-)$ /$\Gamma(\eta^\prime \to \eta \pi^+ \pi^-)$.
{\it Phys Lett B} 2006; {\bf 643}: 41-45.

\bibitem{Ablikim:2016frj}
Ablikim M, Achasov MN and Ahmed S {\em et~al.} (\mbox{BESIII} Collaboration),
Amplitude analysis of the decays $\eta^\prime \rightarrow \pi^+\pi^-\pi^0$ and $\eta^\prime \rightarrow \pi^0\pi^0\pi^0$.
{\it Phys Rev Lett } 2017; {\bf 118}: 012001.

\bibitem{Ablikim:2017kp}
Ablikim M, Achasov MN and Ahmed S {\it et al.} (\mbox{BESIII} Collaboration). 
Measurement of the matrix elements for the decays  $\eta^{\prime}\rightarrow\eta\pi^+\pi^-$ and $\eta^{\prime}\rightarrow\eta\pi^0\pi^0$.
{\it Phys Rev D} 2018; {\bf 97}: 012003.



\bibitem{Borasoy:2005du}
Borasoy B and Nissler R. 
Hadronic $\eta$ and $\eta^\prime$ decays.
{\it Eur Phys J A } 2005; {\bf 26}: 383-398.

\bibitem{Adlarson:2017wlz}
Adlarson P , Afzal F and Ahmed Z {\em et~al.} (A2 Collaboration), 
Measurement of the decay $\eta^{\prime}\to\pi^{0}\pi^{0}\eta$ at MAMI.
{\it Phys. Rev. D} 2018;  \textbf{98}: 012001.


\bibitem{Kubis:2009sb}
Kubis B and Schneider SP.
The cusp effect in $\eta^\prime \to \eta \pi \pi$ decays.
{\it Eur Phys J C} 2009; {\bf 62}: 511-523.

\bibitem{Isken:2017dkw}

Isken T, Kubis B and Schneider SP {\em et al.}
Dispersion relations for $\eta '\rightarrow \eta \pi \pi$.
{\it Eur Phys J  C} 2017; {\bf 77}:  489.









\bibitem{box-anomaly} 
Wess J and Zumino B. Consequences of anomalous ward identities. 
{\it Phys. Lett. B} 1971; {\bf 37}: 95-97.

\bibitem{box-anomaly2}
Witten E.  Global aspects of current algebra.
{\it Nucl. Phys. B} 1983; {\bf 223}: 422-432.

\bibitem{Althoff:1984jq}
Althoff  M, Braunschweig W and Kirschfink FJ {\em et~al.} (TASSO Collaboration).
Measurement of the radiative width of the $\eta^\prime$(958) in two photon interactions.
{\it Phys  Lett B} 1984; {\bf 147}: 487-492.

\bibitem{Aihara:1986sp}
Aihara H, Alston-Garnjost M  and Avery RE
 {\em et~al.} (TPC/Two Gamma Collaboration).
A study of $\eta^\prime$
 formation in photon-photon collisions. 
{\it Phys Rev D} 1987; {\bf 35}: 2650-2654.

\bibitem{Albrecht:1987ed}
Albrecht H,  Andam AA, Binder U {\em et~al.} (ARGUS Collaboration). 
Measurement of $\eta^\prime\to\gamma\pi^+\pi^-$ in $\gamma\gamma$ collisions. 
 {\it Phys Lett B} 1987; {\bf 199}: 457-461.
 
\bibitem{Bityukov:1990db}
Bityukov SI, Borisov GV and Dorofeev VA  {\em et~al.}
Study of the radiative decay $\eta^\prime\to\pi^+\pi^-\gamma$.
{\it Z Phys C} 1991; {\bf 50}: 451-454.

\bibitem{Benayoun:1992ty}
Benayoun M, Feindt M and Girone M {\em et~al.}
Experimental evidences for the box anomaly in $\eta/\eta^\prime$ decays and the electric charge of quarks. 
{\it Z Phys C} 1993; {\bf 58}: 31-54.

\bibitem{Abele:1997yi}
Abele A,  Adomeit J, Amsler C {\em et~al.} (Crystal Barrel Collaboration).
Measurement of the decay distribution of $\eta^\prime\to \pi^+\pi^-\gamma$ and evidence for the box anomaly 
{\it Phys Lett B} 1997;  {\bf 402}: 195-206.
 
 

\bibitem{Ablikim:2017fll}
Ablikim M, Achasov MN and Ahmed S {\em et~al.} (\mbox{BESIII} Collaboration).
Precision study of $\eta^\prime\rightarrow\gamma\pi^+\pi^-$ decay dynamics.
{\it Phys Rev Lett} 2018; {\bf 120}: 242003.

\bibitem{Stollenwerk:2011zz}

Stollenwerk F, Hanhart C and Kupsc A {\em et al.} 
Model-independent approach to $\eta\to\pi^+\pi^-\gamma$ and $\eta^\prime\to\pi^+\pi^-\gamma$.
{\it Phys Lett B} 2012; {\bf 707}: 184-190.






\bibitem{pdg2000}
Zyla RA, Barnett RM and Beringer J {\it et al.} (Particle Data Group). Review of Particle Physics.  {\it Prog Theor Exp Phys} 2020, 083C01. 
\bibitem{DiMicco:2005stk} 
  Di Micco B, Ambrosino F, Antonelli A {\it et al.} (KLOE Collaboration).
  The $\eta \to \pi^0 \gamma \gamma$, $\eta / \eta^\prime$ mixing angle and the $\eta$ mass measurement at KLOE.
 {\it Acta Phys Slov} 2006;  {\bf 56}: 403-409.

\bibitem{Nefkens:2014zlt} 
  Nefkens BMK,  Prakhov S and Aguar-Bartolome P {\it et al.} (A2  Collaboration).
  New measurement of the rare decay $\eta \to \pi^0\gamma\gamma$ with the Crystal Ball/TAPS detectors at the Mainz Microtron.
 {\it Phys Rev C} 2014; {\bf 90}: 025206.

\bibitem{Ablikim:2016tuo}
Ablikim M, Achasov MN and Ahmed S {\em et~al.} (\mbox{BESIII} Collaboration).
Observation of the doubly radiative decay $\eta^{\prime}\to \gamma\gamma\pi^0$.
{\it Phys Rev D} 2017; {\bf 96}: 012005.

\bibitem{Ablikim:2019wsb}
Ablikim M, Achasov MN,  Adlarson P {\em et~al.} (\mbox{BESIII} Collaboration).
Search for $\eta^\prime\to\gamma\gamma\eta$.
{\it  Phys Rev D} 2019; {\bf 100}: 052015.


\bibitem{Escribano:2018cwg}

Escribano R,  Gonzàlez-Solís S and Jora R {\em et al.}
A theoretical analysis of the doubly radiative decays $\eta^{(\prime)}\to\gamma\gamma\pi^0$ and $\eta^\prime\to\gamma\gamma\eta$.
{\it Phys Rev D} 2020; {\bf 102}: 034026.



\bibitem{Landsberg:1986fd}
Landsberg LG. 
Electromagnetic decays of light mesons.
{\it Phys Rept } 1985; {\bf 128}: 301-376.

\bibitem{Aoyama:2020ynm}

Aoyama T, Asmussen N and Benayoun M  \textit{et al.}
The anomalous magnetic moment of the muon in the Standard Model.
{\it Phys. Rept.}  2020; \textbf{887} :1-166.



\bibitem{snd}

Achasov MN, Aulchenko VM and Beloborodov KI
{\em et~al.} (SND Collaboration). 
Study of conversion decays $\phi\to\eta e^+e^-$ and $\eta \to \gamma e^+ e^-$ in the experiment with SND Detector at the VEPP-2M collider. 
{\it Phys Lett B} 2001; {\bf  504}: 275-281.

\bibitem{wasa}

Berlowski M, Bargholtz Chr. and  Bashkanov M {\em et~al.}
 (CELSIUS/WASA Collaboration).  Measurement of eta meson decays into lepton-antilepton pairs.
{\it Phys Rev D} 2008; {\bf  77}:  032004.

\bibitem{Ablikim:2015wnx}
Ablikim M,  Achasov MN and Ai XC {\em et~al.} (\mbox{BESIII} Collaboration).
Observation of the Dalitz decay $\eta' \to \gamma e^+e^-$.
{\it Phys Rev D} 2015; {\bf 92}: 012001.



\bibitem{Bramon:1981sw}
Bramon A and Masso E.
$Q^2$ Duality for electromagnetic form-factors of mesons. 
{\it Phys Lett B} 1981;  {\bf 104}: 311-314.

\bibitem{Ametller:1983ec}

Ametller L, Bergstrom L and Bramon A {\it et al.}
The quark triangle: application to pion and $\eta$ decays.
{\it Nucl Phys B} 1983; {\bf 228}: 301-315.


\bibitem{Ametller:1991jv}

Ametller L, Bijnens J and Bramon A {\it et al.}
Transition form-factors in $\pi^0$, $\eta$ and $\eta^\prime$ couplings to gamma gamma. 
{\it Phys Rev D} 1992;  {\bf 45}: 986.

\bibitem{Hanhart:2013vba}

Hanhart C, Kupsc A and Mei{\ss}ner UG {\it et al.}
Dispersive analysis for $\eta\to\gamma\gamma^*$.
{\it Eur Phys J C} 2013; {\bf 73}: 2668. Erratum: {\it Eur Phys J C} 2015; {\bf 75}: 242.

\bibitem{Dzhelyadin:1979za}

Dzhelyadin RI, Golovkin SV and Gritzuk MV 
 {\em et~al.}
Observation of $\eta^\prime\to \mu^+\mu^-\gamma$ decay. 
{\it Phys Lett  B} 1979; {\bf 88}: 379-380.


\bibitem{KLOE2:2011aa}
Ambrosino F, Antonelli A, Antonelli M {\it et al.} (KLOE-2 Collaboration).
Observation of the rare $\eta\to e^+e^-e^+e^-$ decay with the KLOE experiment.
{\it Phys Lett B} 2011; {\bf 702}; 324. 
\bibitem{Petri:2010ea}
Petri T, Anomalous decays of pseudoscalar mesons.
 arXiv:1010.2378.
  
 \bibitem{Escribano:2015vjz}

Escribano R and Gonz\`alez-Sol\'\i{}s S.
A data-driven approach to $\pi^{0}, \eta$ and $\eta^{\prime}$ single and double Dalitz decays.
{\it Chin. Phys. C} 2018;  {\bf 42}: 023109.
  
  \bibitem{Czerwinski:2012ry} Czerwinsk  E and Eidelman S and Hanhart C {\em et al.} 
  MesonNet workshop on meson Transition Form Factors.
arXiv:1207.6556.








\bibitem{Niecknig:2012sj} Niecknig F, Kubis B and Schneider SP.
Dispersive analysis of $\omega\to 3\pi$ and $\phi\to 3\pi$.
{\it Eur Phys J C} 2012; {\bf 72}:2014.  

\bibitem{Danilkin:2014cra} Danilkin IV, Fernandez-Ramirez C and Guo P {\it et al.}
Dispersive analysis of $\omega/\phi\to 3\pi,\pi\gamma^*$.
{\it  Phys Rev D } 2015; {\bf 91}: 094029.


\bibitem{Ablikim:2018yen}
Ablikim M, Achasov MN, Ahmed S {\em et~al.} (\mbox{BESIII} Collaboration).
Dalitz plot analysis of the decay $\omega\to\pi^+\pi^-\pi^0$.
{\it Phys Rev  D} 2018; {\bf 98}:112007.


  \bibitem{Adlarson:2016wkw}%

  Adlarson P, Augustyniakb W and Bardanc W 
   {\em et~al.} (WASA-at-COSY Collaboration).
  Measurement of the $\omega\to\pi^+\pi^-\pi^0$ Dalitz plot distribution.
  {\it Phys Lett B } 2017; {\bf 770}; 418-425. 
  
  




  
    
  
  \bibitem{Chen:2007xr}
  Chen HX,  Hosaka A and Zhu SL.  Light scalar tetraquark mesons in the QCD Sum Rule.
{\it  Phys Rev D} 2007; {\bf 76}: 094025.



\bibitem{RuizdeElvira:2010cs}

  Ruiz de Elvira J, Pelaez JR and Pennington MR {\it et al.}  Chiral Perturbation Theory, the ${1/N_c}$ expansion and Regge behaviour determine the structure of the lightest scalar meson.
 {\it Phys Rev D} 2011; {\bf 84}: 096006.
 


\bibitem{Hooft:2008we}
't Hooft G, Isidori G and Maiani L {\it et al.} A theory of scalar mesons.
{\it Phys  Lett B} 2008; {\bf 662}: 424-430.



  \bibitem{Parganlija:2010fz}
  Parganlija D, Giacosa  F and Rischke DH. 
  Vacuum properties of mesons in a Linear Sigma Model with vector mesons and global chiral invariance.
  {\it Phys Rev D} 2010; {\bf 82}: 054024.
  
    \bibitem{Ablikim:2004qna}  Ablikim M, Bai JZ and Ban Y {\it et al.} (BES Collaboration).
    The $\sigma$ ploe in $J/\psi\to\omega\pi^+\pi^-$.
    {\it Phys Lett B} 2004; {\bf  598}: 149-158.
    
  
  \bibitem{beskappa}  Ablikim M, Bai JZ and Ban Y {\it et al.} (BES Collaboration).
  Evidence for $\kappa$ meson production in  $J/\psi\to {K^*(892)^0}K^+\pi^-$ process.
    {\it Phys Lett B} 2006; {\bf  633}: 681-690.
  
  \bibitem{besf0} Ablikim M, Bai JZ and Ban Y {\it et al.} (BES Collaboration).
  Resonances in $J/\psi\to\phi\pi^+\pi^-$ and $\phi K^+K^-$.
  {\it Phys. Lett B} 2005; {\bf 607}: 243-253.
  
  \bibitem{Ablikim:2018pik}
    Ablikim M, Achasov MN and Adlarson P {\it et al.}   (\mbox{BESIII} Collaboration).
Observation of $a_{0}(980)$-$f_{0}(980)$ Mixing.
 {\it Phys Rev Lett} 2018; {\bf 121}: 022001.
 


    \bibitem{Ablikim:2019pit}
    Ablikim M, Achasov MN and Adlarson P {\it et al.}   (\mbox{BESIII} Collaboration).
    Amplitude analysis of $D_{s}^{+} \rightarrow\pi^{+}\pi^{0}\eta$ and first observation of the pure $W$-annihilation decays $D_{s}^{+} \rightarrow
 a_{0}(980)^{+}\pi^{0}$ and $D_{s}^{+} \rightarrow a_{0}(980)^{0}\pi^{+}$.
    {\it Phys Rev Lett } 2019;  {\bf 123}: 112001.
      



\bibitem{Gan:2020aco}

Gan L, Kubis B and Passemar E {\it et al.}
Precision tests of fundamental physics with $\eta$ and $\eta^\prime$ mesons.
arXiv:2007.00664 [hep-ph].


\bibitem{Jarlskog:2002zz}
 Jarlskog  C and Shabalin E.
On searches for CP, T, CPT and C violation in flavour-changing and flavour-conserving interactions.
{\it Phys Scripta  T} 2002; {\bf 99}: 23-33. 

\bibitem{Jarlskog:1995gz}
Jarlskog C and  Shabalin E.
$\epsilon^\prime$ and the decay $\eta\to \pi\pi$ in a theory
                        with both explicit and spontaneous CP violation.
{\it  Phys Rev D} 1995; {\bf 52}: 6327-6335.



\bibitem{Ablikim:2011vg}
Ablikim M, Achasov MN and  Alberto D {\em et~al.} (\mbox{BESIII} Collaboration).
Search for CP and $P$ violating pseudoscalar decays into $\pi\pi$.
{\it Phys Rev D} 2011; {\bf 84}: 032006.



\bibitem{Ablikim:2019msz}
Ablikim M, Achasov MN and Adlarson P {\em et~al.} (\mbox{BESIII} Collaboration ).
Search for the rare decay $\eta'\rightarrow\pi^{0}\pi^{0}\pi^{0}\pi^{0}$ at \mbox{BESIII}.
{\it Phys Rev D} 2019;{\bf 101}: 032001.


\bibitem{Geng:2002ua}

Geng CQ, Ng JN and Wu TH. 	
CP violation in the decay $\eta\to\pi^+\pi^-\gamma$. 
 {\it Mod Phys Lett A} 2002;  {\bf 17}:
  1489-1498. 
  
  \bibitem{Ambrosino:2008cp}
Ambrosino F, Antonelli  A and Antonelli M {\em et~al.} (KLOE Collaboration).
Measurement of the branching ratio and search for a CP violating asymmetry in the $\eta \to \pi^+ \pi^- e^+e^-(\gamma)$ decay at KLOE.
{\it Phys Lett  B} 2009;  {\bf 675}: 283-288. 

\bibitem{Ablikim:2013wfg}
Ablikim M, Achasov MN and Adlarson P  {\em et~al.} (\mbox{BESIII} Collaboration).
Measurement of $\eta^\prime\rightarrow \pi^+\pi^-e^+e^-$ and $\eta^\prime\rightarrow \pi^+\pi^-\mu^+\mu^-$
{\it  Phys Rev D}  2013; {\bf 87}: 092011.

\bibitem{Ablikim:2020svz}

Ablikim M, Achasov MN and Adlarson P  {\em et~al.} (\mbox{BESIII} Collaboration). 
Observation of $\eta^\prime\rightarrow\pi^+\pi^-\mu^+\mu^-$.
arXiv: 2012.04257[hep-x]. 

\bibitem{Escribano:2020rfs}
 Escribano R and Royo E.
A theoretical analysis of the semileptonic decays $\eta^{(\prime)}\to\pi^0l^+l^-$ and $\eta^\prime\to\eta l^+l^-$.
 {\it Eur. Phys. J. C} 2020; {\bf 80}: 1190.

  
  \bibitem{scattering}
  Druzhinin VP,   Eidelman SI and  Serednyakov SI {\it et al.} 
  Hadron production via $e^+e^-$ collisions with Initial State Radiation. 
    {\it Rev Mod Phys} 2011; {\bf 83}: 1545-1588. 
  
  \bibitem{babar}

  Aubert B, Bona M and  Boutigny D {\em et al.} (BaBar Collaboration).
The $e^+e^- \to K^+ K^- \pi^+ \pi^-, K^+ K^- \pi^0 \pi^0$ and $K^+ K^- K^+ K^-$ cross-sections measured with initial-state radiation.
{\it Phys. Rev. D} 2007; {\bf 76} : 012008. 


  \bibitem{bes}

  Ablikim M, Bai JZ and Bai Y {\it et al.} (BES Collaboration).
  Observation of $Y(2175)$ in $J/\psi \to  \eta\phi f_0(980)$. 
 {\it Phys. Rev. Lett.} 2008; {\bf 100}: 102003.
  
  
  \bibitem{besKK} 

  Ablikim M, Achasov MN and Adlarson P  {\em et~al.} (\mbox{BESIII} Collaboration).
  Measurement of $e^+e^-\to K^+K^-$ cross section at $\sqrt{s}=$ 2.00-3.08 GeV.
  {\it Phys Rev D} 2019; {\bf 99}: 032001.
  
  \bibitem{KKpi0pi0} Ablikim M, Achasov MN and Adlarson P  {\it et al.} (\mbox{BESIII} Collaboration).
  Observation of a resonant structure in $e^{+}e^{-} \to K^{+}K^{-}\pi^{0}\pi^{0}$.
  {\it Phys Rev Lett} 2020;  {\bf 124}:112001.
  
    \bibitem{Ablikim:2020coo} Ablikim M, Achasov MN and Adlarson P  {\it et al.} (\mbox{BESIII} Collaboration).
    Observation of a structure in $e^{+}e^{-} \to \phi\eta^{\prime}$ at $\sqrt{s}$ from 2.05 to 3.08 GeV
    {\it Phys Rev D} 2020; {\bf 102}: 012008.

  \bibitem{Ablikim:2014pfc} Ablikim M, Achasov MN and Adlarson P {\it et al.} (\mbox{BESIII} Collaboration).
  Study of $J/\psi \to \eta \phi \pi^+ \pi^-$ at \mbox{BESIII}. 
  {\it Phys Rev D} 2015; {\bf 91}: 052017.

\bibitem{KKKK} Ablikim M, Achasov MN  and Adlarson P {\it et al.} (\mbox{BESIII} Collaboration).
  Cross section measurements of $e^{+}e^{-} \to
                        K^{+}K^{-}K^{+}K^{-} $ and $ \phi K^{+}K^{-}$ at
                        center-of-mass energies from 2.10 to 3.08 GeV.      
{\it  Phys Rev D} 2019; {\bf 100}: 032009.
        
  
 \bibitem{focus} 
 Link JM, Reyes M and Yager PM
 {\it et al.} (FOCUS Collaboration).
  Observation of a 1750 MeV/c$^2$ enhancement in the diffractive photoproduction of $K^+K^-$.
  {\it Phys. Lett. B} 2002; {\bf 545}: 50-56.
 \bibitem{Ablikim:2019xhx} Ablikim M, Achasov MN and Adlarson P {\it et al.} (\mbox{BESIII} Collaboration).
 Partial wave analysis of $\psi(3686)\to K^+K^-\eta$.    
     {\it Phys Rev D} 2010; {\bf 101}: 032008.
   
  
  
\end{thebibliography}
\end{document}